\pdfoutput=1 
\documentclass[pre,twocolumn,showpacs,amsmath,amssymb,floatfix,eqsecnum]{revtex4-1}

\usepackage{graphicx}
\usepackage{times}

\begin{document}

\def \beq {\begin{equation}}
\def \eeq {\end{equation}}
\def \beqa {\begin{eqnarray}}
\def \eeqa {\end{eqnarray}}
\def \al {\alpha}
\def \mb {\mathbf}
\def \mbs {\boldsymbol}
\def \nnb{\nonumber}
\def \A {\mathcal{A}}
\def \R {\mathcal{R}}
\def \th {\theta}
\def \sg {\sigma}
\def \ep {\epsilon}
\def \vphi {\varphi}
\def \lra {\leftrightarrow}
\def \pr {\partial}
\def \pz {\partial_z}
\def \pth {\partial_{\theta}}
\def \psg {\partial_{\sg}}
\newcommand{\tab}{\hspace*{2em}}

\title{Swimming at Low Reynolds Number in Fluids with Odd (Hall) Viscosity}

\author{Matthew F. Lapa and Taylor L. Hughes}
\affiliation{Department of Physics and Institute for Condensed Matter Theory, University of Illinois at Urbana-Champaign, 61801-3080}

\date{\today}

\begin{abstract}

We apply the geometric theory of swimming at low Reynolds number to the study of nearly circular swimmers in
two-dimensional fluids with non-vanishing Hall, or ``odd", viscosity. The Hall viscosity gives an off-diagonal contribution
to the fluid stress-tensor, which results in a number of striking effects. In particular, we find that a swimmer whose area
is changing will experience a torque proportional to the rate of change of the area, with the constant of proportionality 
given by the coefficient $\eta^o$ of odd viscosity. After working out the general theory of swimming in fluids with 
Hall viscosity for a class of simple swimmers, we give a number of example swimming strokes which clearly demonstrate
the differences between swimming in a fluid with conventional viscosity and a fluid which also has a Hall viscosity. A number of 
more technical results, including a proof of the torque-area relation for swimmers of more general shape, are explained in a set 
of appendices.

\end{abstract}


\maketitle

The theory of swimming in classical fluids at low Reynolds number \cite{Taylor1951,Purcell1976} is remarkable because of the 
connections it makes between seemingly disparate fields \cite{Wilc1987}. For example, the motion of swimmers with cyclic  
swimming strokes is determined purely from classical fluid dynamics, but it can be re-cast into an elegant geometric formulation 
reminiscent of Berry's phase physics and gauge fields \cite{Wilc1987,Wilc1989,Wilc1989-3}. In fact, the motion of tiny organisms 
in fluids with high viscosity can be captured by a ``gauge-theory" of shapes. Since the initial work on the geometric formulation 
of swimming there have been generalizations to swimmers in quantum fluids \cite{avron2006quantum} and even to swimmers in 
fluids on curved spaces \cite{Wisdom2003,Avr2006}. The theory has also been successfully applied in practice to describe the 
swimming of robots \cite{Becker2003} and microbots \cite{najafi2004,dreyfus2005}.

In this article we focus on swimmers in 2D fluids with broken time-reversal symmetry, for example, fluids in magnetic fields or 
rotating fluids. We are not interested in the specific source of time-reversal breaking, but instead just consider a classical fluid with a 
microscopic source of local angular momentum (on a much smaller scale than the size of the swimmer) that gives rise to a 
non-vanishing Hall (or odd) viscosity coefficient \cite{Avr1995,Avr1998} in addition to the usual isotropic viscosity coefficients. The Hall 
viscosity is an off-diagonal viscosity term that is dissipationless and produces forces perpendicular to the direction of the fluid 
flow. It can have a quantum mechanical origin in, for example, systems exhibiting the quantum Hall 
effect 
\cite{Avr1995,levay1997,Avr1998,Read2009,tokatly2009,haldane2009,Hughes2011,read2011,bradlyn2012,Son2012,hughes2013},
 or a classical origin in plasmas at finite-temperature \cite{LandauKinetics}. 

We will not focus on the microscopic origin of the Hall viscosity coefficient, but only assume it to be non-vanishing in conjunction with the usual viscosity coefficients. From this 
assumption we will determine the motion of swimmers at low Reynolds number in the presence of Hall viscosity. Specifically, we will 
consider the problem of swimmers with circular boundaries that move via deformations of their boundaries analogous to the 
nearly-circular swimmers in Ref. \cite{Wilc1987,Wilc1989}. We find a general result that connects the torque on a swimmer to the 
rate of area change of the swimmer with a proportionality constant given by the Hall viscosity. We use our results to give 
examples of swimmer motion due to cyclic circular deformations and compare cases where the conventional and Hall viscosities 
each dominate. Our paper is organized as follows: we first review the geometric formulation of swimming and the appearance of 
Hall viscosity in 2D fluids with broken time-reversal symmetry. We then go on to derive the general consequences of the Hall 
viscosity on swimmers and then give explicit examples of model swimming strokes that illustrate some differences between fluids 
with vanishing and non-vanishing Hall viscosity. Finally, we have some appendices which collect derivations of the more 
technical results. Note that we will use Hall viscosity and odd viscosity interchangeably throughout the text. 

\section{Review of Geometric Formulation of the Swimming Problem}
We begin by reviewing the geometric formulation of the problem of swimming at low Reynolds number developed by Shapere
and Wilczek \cite{Wilc1987,Wilc1989}. The instantaneous rigid motion (translation and rotation) 
of a swimmer is determined by the condition that
the swimmer not be able to exert a net force or torque on itself, and the condition
 that the fluid velocity vanishes at infinity.

We should first explain why the problem of swimming at low Reynolds number can be formulated in a purely geometric
way, independent of the mass of the swimmer or the speed of the swimming stroke (assuming the speed of the 
stroke is still small enough so that there is no appreciable momentum transfer to the fluid).
Recall that the Reynolds number, which is associated with a viscous fluid and an object in motion in that fluid, is a 
ratio of the inertial and viscous
forces on that object (we are not yet considering systems with odd viscosity so in this sentence the word ``viscous" 
refers to the traditional dissipative (even) viscosity of the fluid). If $\eta^e$ is the even viscosity coefficient, $V$ is a
typical speed of the fluid flow, $L$ is a characteristic dimension of the swimming object, and $\rho$ is the density of the fluid, then
the Reynolds number can be expressed as
\beq
	\text{Re} = \frac{\rho VL}{\eta^e}\ .
\eeq 
The low Reynolds number regime can be interpreted as the regime where the momentum density of the fluid, $\rho V$, 
is negligible compared to the scale $\eta^e/L$.

At low Reynolds number the drag force on the swimmer is proportional to its velocity. 
This means that if the swimmer stops its stroke
and just coasts through the fluid, its speed will decay exponentially until it comes to a stop. In the low Reynolds number regime this exponential decay is so fast
that the motion of the swimmer at any given time can be considered to be completely independent of
 what the swimmer was doing at all previous times \cite{Purcell1976}. 
The motion of the swimmer at time $t$ depends only on its shape and the velocity of its surface at time $t$.
With these remarks in mind we can move on to discuss the geometric theory of swimming at low Reynolds number.

In two dimensions, for swimmers modeled as the interior of deformed circles,  we can represent the swimming stroke (the motion of the boundary of the swimmer) by a time-dependent
 complex function $S_0(\sg,t)$, $\sg = e^{i\th}$, whose real and imaginary parts give the $x$ and $y$ positions of the 
point on the swimmer described by the parameter $\theta \in [0,2\pi)$ at the time $t$. When we want to emphasize the 
dependence of $S_0(\sg,t)$ on the real parameter $\th$ instead of the complex parameter $\sg$
(as we do in Appendix \ref{sec:AreaProof}) we call it $S_0(\th,t)$ instead.

The function $S_0(\sg,t)$ lives in a space of ``un-located" shapes, which can be
obtained from the space of ``located" shapes by partitioning it into equivalence classes $[S_0(\sg,t)]$
containing all shapes differing only by a rigid motion. The location and orientation of the swimmer in real space 
is specified by a rigid motion $\R(t)$ acting on a representative of the equivalence class $[S_0(\sg,t)]$, the simplest
choice being $S_0(\sg,t)$ itself:
\beq
	S(\sg,t) = \R(t)S_0(\sg,t)\ .
\eeq
To take an example, $S_0(\sg,t)$ might be the representative of $[S_0(\sg,t)]$ with its centroid at the origin and
a distinguishing feature of the shape aligned with the x-axis at time $t=0$. 

To be concrete, let us encode the translation and rotation represented by $\R(t)$ into a $3\times3$ matrix and let this matrix
act  on $S_0(\sg,t)$ represented as a three-dimensional vector with third entry equal to one,
\beq
	\R(t)S_0(\theta,t) = \begin{pmatrix}
		\cos(\Theta) & \sin(\Theta) & X \\
		-\sin(\Theta) & \cos(\Theta) & Y \\
		0 & 0 & 1
	\end{pmatrix} \begin{pmatrix}
		\text{Re}[S_0(\sg,t)] \\
		\text{Im}[S_0(\sg,t)] \\
		1
	\end{pmatrix}\ ,
\eeq
where $(X,Y)$ and $\Theta$ are the vector and angle representing the translation and rotation effected by $R(t)$.
The matrix $\R(t)$ is determined by integrating the equation
\beq
	\frac{d \R(t)}{dt} = \R(t)\A(t)\ ,\label{eq:diffeq}
\eeq
where the matrix $\A(t)$ determines the infinitesimal rigid motion of the swimmer during a time $dt$ in the sense that
$\A(t)\ dt$ \emph{is} the rigid motion of the swimmer during the interval $dt$. 
The matrix $\A(t)$ is completely determined by the
requirements that the net force and torque on the swimmer vanish and that the fluid velocity goes to zero at infinity. To 
determine the swimming path we need to find $\A (t)$ for a given swimming stroke and then integrate Eq. \ref{eq:diffeq}.

Integrating this equation gives the solution for the rigid motion $\R(t)$,
\beq
	\R(t) = \R(0) \bar{P} \text{exp}\left[ \int_0^t \A(t_1) dt_1 \right]\ ,
\eeq
where $\bar{P}$ denotes a reverse path-ordering operation. Explicitly, we have
\beqa
	\bar{P} \text{exp}\left[ \int_0^t \A(t) dt \right] &=& I + \int_0^t \A(t) dt \nnb \\
 &+& \int_0^t \left( \int_0^{t_1} \A(t_2)\A(t_1) dt_2 \right) dt_1 \nnb \\
 &+& \dots \label{POI}
\eeqa
so the matrix $\A(t_i)$ with the latest time $t_i$ appears furthest to the right in each integral, which is the reverse
of the usual path ordering operation where the latest time goes furthest to the left in each integral. We show how this 
integration is carried out numerically in Appendix \ref{app:computation}.

To see how the idea of a gauge theory of shapes enters we first note that the choice of a representative from the equivalence 
class $[S_0(\sg,t)]$ is analogous
to a choice of gauge, and the matrix $\A(t)$ plays the role of a gauge potential. If we choose a different representative
$\tilde{S}_0(\sg,t)$, related to $S_0(\sg,t)$ by a rigid motion $U(t)$ (we can choose a different representative at
each time t),
\beq
	\tilde{S}_0(\sg,t) = U(t) S_0(\sg,t)\ ,
\eeq
then the requirement that the rigid motion of the swimmer in real space remain unchanged leads to the transformation
law for $\R(t)$
\beq
	\R(t) \to \R'(t) = \R(t)U^{-1}(t)\ .
\eeq
The fact that the transformed gauge potential must satisfy the new differential equation
\beq
	\frac{d \R'(t)}{dt} = \R'(t)\A'(t)
\eeq
yields the familiar transformation law
\beq
	\A(t) \to \A'(t) = U(t)\A(t)U^{-1}(t) + U(t)\frac{d U^{-1}(t)}{dt}\ ,
\eeq
which shows that $\A(t)$ does indeed transform like a gauge potential.

We can also represent $\A(t)$ in the form of a $3\times3$ matrix,
\beq
	\A(t) =  \begin{pmatrix}
		0 & \omega & V_x \\
		-\omega & 0 & V_y \\
		0 & 0 & 0
	\end{pmatrix} \label{gauge}
\eeq
where $(V_x,V_y)$ and $\omega$ are the instantaneous linear and angular velocity of the swimmer (so $\A(t)$ is
in the Lie Algebra of rigid motions). Sometimes we will refer to the translational and rotational parts $\A_{tr}$ and $\A_{rot}$
of the gauge potential, defined by
\begin{subequations}
\beqa
	\A_{tr} &=& V_x + iV_y \\
	\A_{rot} &=& \omega\ .
\eeqa
\end{subequations}
The components of $\A(t)$ can be completely determined by solving the equations
of motion for Stokes flow of the viscous fluid surrounding the swimmer, subject to no-slip boundary conditions at the
surface of the swimmer. Now that we have reviewed the geometric formulation of swimming we will introduce the concept of the odd/Hall viscosity in time-reversal breaking fluids.

\section{Odd Viscosity}
\label{sec:OddVisc}

We now review the basic definition of odd viscosity and the derivation of the isotropic odd viscosity contribution
to the fluid stress tensor in two dimensions. Throughout this section we follow the presentation of Ref.~\onlinecite{Avr1998} 
where most of these details were first worked out. 

The general linear relation between the fluid stress tensor $T_{ij}$ and the rate of strain tensor 
$v_{ij} = \tfrac{1}{2}(\partial_j v_i + \partial_i v_j)$ ($v_i$ are the components of the fluid velocity vector $\mb{v}$)
 is of the form 
\beq
	T_{ij} = \eta_{ijkl} v_{kl}\ .
\eeq
The symmetry of the stress and rate of strain tensors imply the symmetry of the viscosity tensor
$\eta_{ijkl}$ under the exchanges
$i \lra j$ and $k \lra l$, but in general $\eta_{ijkl}$ can contain terms which are symmetric or anti-symmetric under
the exchange of the pair of indices $\{ij\}$ with the pair of indices $\{kl\}$. We can always split $\eta_{ijkl}$ into parts
which are even and odd under such an exchange by writing $\eta_{ijkl} = \eta^e_{ijkl} + \eta^o_{ijkl}$.

To extract the isotropic contribution to $\eta^o_{ijkl}$ it is convenient to use a simple basis for representing a real, 4th-rank 
tensor that is symmetric under exchange of its first two and second two indices. One such basis is provided by the tensor 
products
\beq
	\sg^a \otimes \sg^b,\ a,b \in \{0,1,3\}
\eeq
of the Pauli matrices $\sg^1$, $\sg^3$ and the $2\times 2$ identity matrix $\sg^0$, where we have been careful to only use 
the symmetric matrices. We can expand the viscosity tensor as 
\beq
	\eta_{ijkl} = \sum_{a,b = 0, 1,3} \eta_{ab} \sg^a_{ij}\sg^b_{kl}
\eeq
and then identify the odd part as
\beq
	\eta^o_{ijkl} = \sum_{a \neq b} \eta^o_{ab} ( \sg^a_{ij}\sg^b_{kl} - \sg^b_{ij}\sg^a_{kl} )\ .
\eeq

In two dimensions the generator of spatial rotations is $i\sg^2$, where $\sg^2$ is the second Pauli matrix. In an isotropic fluid
the viscosity tensor must commute with $\sg^2\otimes\sg^2$ to be rotationally invariant. Using the familiar commutation and anti-commutation relations
for the Pauli matrices, and the fact that all matrices commute with the identity $\sg^0$, we find that in an isotropic fluid the odd part of
the viscosity tensor must have the form
\beq
	\eta^o_{ijkl} = \eta^o ( \sg^1_{ij}\sg^3_{kl} - \sg^3_{ij}\sg^1_{kl} )\ ,
\eeq
where the single constant $\eta^o$ is the coefficient of odd viscosity. Finally we can use the explicit expressions
\begin{subequations}
\beqa
	\sg^1_{ij} &=& \delta_{i1}\delta_{j2} + \delta_{i2}\delta_{j1} \\
	\sg^3_{ij} &=& \delta_{i1}\delta_{j1} - \delta_{i2}\delta_{j2}
\eeqa
\end{subequations}
for the elements of the Pauli matrices $\sg^1$ and $\sg^3$ to write down the form of the odd viscosity contribution
to the stress tensor
\beqa
	T^o_{ij} &=& \eta^o_{ijkl}v_{kl} \nnb \\
	&=& - 2\eta^o (\delta_{i1}\delta_{j1} - \delta_{i2}\delta_{j2})v_{12} \nnb \\
	&+& \eta^o (\delta_{i1}\delta_{j2} + \delta_{i2}\delta_{j1})(v_{11} - v_{22}) 
\eeqa\noindent which was first obtained in Ref. \onlinecite{Avr1998}.
For comparison we also display the much more familiar even viscosity part of the stress tensor (for an incompressible 
fluid)
\beq
	T^e_{ij} = 2\eta^{e}v_{ij}\ ,
\eeq
where $\eta^e$ is the coefficient of even viscosity.

\begin{figure}[t]
  \centering
    \includegraphics[width= .5\textwidth]{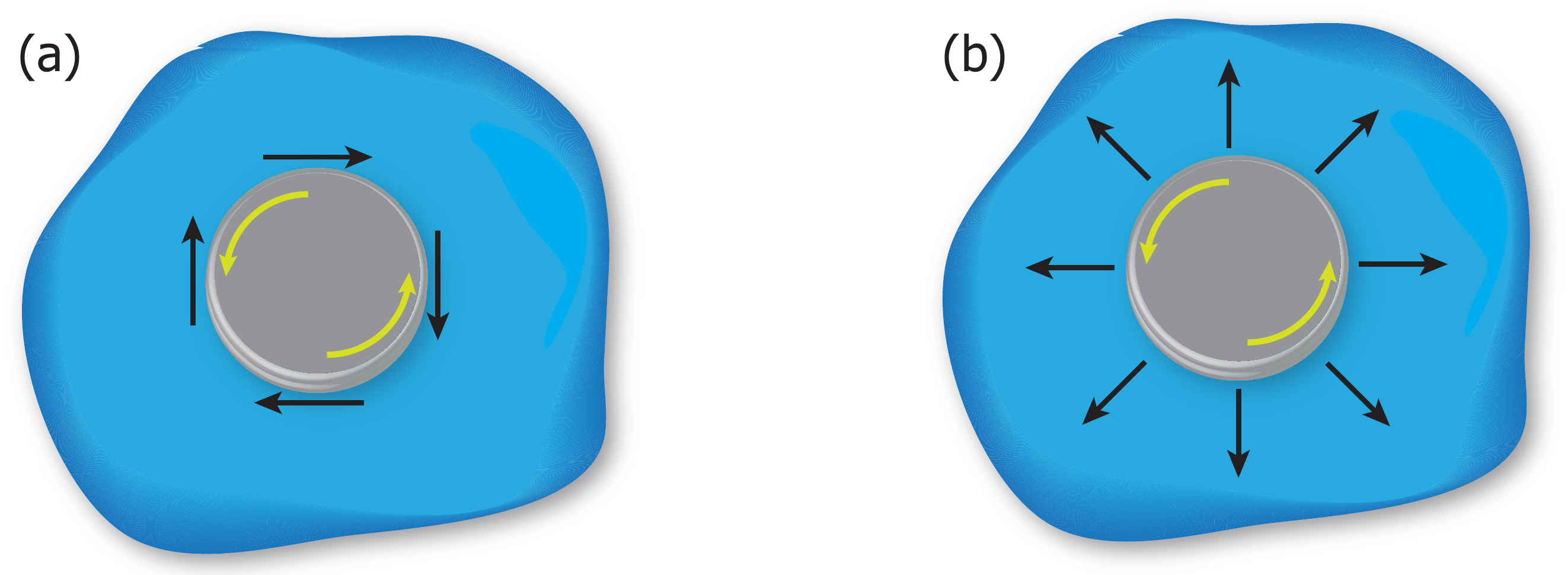} 
\vskip 10pt
 \caption{a) In a fluid with even viscosity only a rotating circle will feel a torque that opposes its rotation and is proportional to 
the coefficient of even viscosity $\eta^e$. b) In a fluid with odd viscosity a rotating circle will also feel a pressure directed 
radially inwards or outwards (depending on the direction of the rotation) and proportional to the coefficient of odd viscosity
$\eta^o$. The dependence of this pressure force on the direction of the rotation indicates that time-reversal symmetry is
broken in systems with non-vanishing odd viscosity.}
\label{fig:droplet}
\end{figure}

We see that diagonal elements of $T^o_{ij}$ are proportional to off-diagonal elements of $v_{ij}$ and off-diagonal
elements of $T^o_{ij}$ are proportional to diagonal elements of $v_{ij}$. This atypical relation between 
the elements of $T^o_{ij}$ and $v_{ij}$ has a number of non-intuitive consequences. For example, a circular
object rotating in a fluid with odd viscosity will feel a pressure, directed either radially inwards or outwards depending
on the sense of the rotation (see \cite{Avr1998}  and Fig. \ref{fig:droplet}). This is quite
different from what would happen in a fluid with even viscosity only, where a rotating circle would feel a torque that
opposes the rotation. The fact that the direction of the pressure force (radially inwards or outwards)
on a circle rotating in an odd viscosity fluid depends on the sense of the rotation means that
time-reversal symmetry is broken in systems with odd viscosity.

\section{Equations of Motion, Force and Torque} 

In classical fluids with both even and odd viscosity Avron has shown (see  \cite{Avr1998}) that the equations
of motion for incompressible Stokes flow (viscous force-dominated flow) are
\begin{subequations}
\beqa
	\nabla(p - \eta^o\xi) &=& \eta^e \nabla^2\mb{v} \\
	\nabla\cdot\mb{v} &=& 0\ ,
\eeqa
\end{subequations}
where $p$ is the pressure, $\xi = (\nabla\times\mb{v})\cdot\mb{\hat{z}}$ is the vorticity, and $\eta^e$ and $\eta^o$ are the
coefficients of even and odd viscosity, respectively. If $\eta^e \neq 0$, taking the curl of the first equation shows that the 
vorticity is a harmonic function, i.e., $\nabla^2 \xi = 0$. This means that the stream function $\psi$ (which can be used here
because the flow is incompressible), defined by $\mb{v} = \nabla\times(\psi\hat{\mb{z}})$, is a biharmonic function,
\beq
	\nabla^2(\nabla^2\psi) = 0\ .
\eeq
In two dimensions we can package the velocity vector $\mb{v}=(v_1,v_2)$ into a complex variable $v = v_1 + iv_2$. 
The solution for $v$ can then be expressed in the complex form (see \cite{Wilc1989})
\beq
	v = \phi_1(z) - z\overline{\pz\phi_1(z)} + \overline{\phi_2(z)} \label{velocity}
\eeq
where $z= x+iy = Re^{i\vphi}$, the bar denotes complex conjugation and $\pz = \tfrac{1}{2}(\pr_1 - i\pr_2)$. 
The functions $\phi_1(z)$ and $\phi_2(z)$ are analytic functions 
(away from the point $z=0$, which lies inside the swimmer) with the Laurent series expansions
\begin{subequations}
\beqa
	\phi_1(z) &=& \sum_{k<0}a_k z^{k+1} \\
	\phi_2(z) &=& \sum_{k<-1}b_k z^{k+1} \ .
\eeqa
\end{subequations}
To solve for the coefficients $a_k$ and $b_k$ we impose no-slip boundary conditions at the surface of the swimmer. 
Solving for these coefficients can be very difficult for general swimming strokes, so we will focus our attention on a class
of simple swimmers introduced in Ref. \onlinecite{Wilc1989} whose shapes are conformal maps of the circle of degree ${\cal{D}}=2$. In 
Appendix \ref{app:Degree3} we extend our results to swimmers that are conformal maps of the circle of degree ${\cal{D}}=3$.

To calculate the force and torque on the swimmer we will need the stress tensor. We have seen in Section 
\ref{sec:OddVisc} that in the presence of odd viscosity the stress tensor gets an extra contribution. 
The full stress tensor is now
\beqa
	T_{ij} &=& -p\delta_{ij} + 2\eta^{e}v_{ij} - 2\eta^o (\delta_{i1}\delta_{j1} - \delta_{i2}\delta_{j2})v_{12}\nnb \\ 
&+& \eta^o (\delta_{i1}\delta_{j2} + \delta_{i2}\delta_{j1})(v_{11} - v_{22})\ .  \label{odd_stress}
\eeqa
The components of the odd-viscosity part of the stress tensor are
\begin{subequations}
\beqa
	T^o_{11} &=& -\eta^o(\partial_2 v_1 + \partial_1 v_2) \\
	T^o_{12} &=& \eta^o(\partial_1 v_1 - \partial_2 v_2) \\
	T^o_{21} &=& \eta^o(\partial_1 v_1 - \partial_2 v_2) \\
	T^o_{22} &=& \eta^o(\partial_2 v_1 + \partial_1 v_2)\ .
\eeqa
\end{subequations}

Since the fluid is incompressible, an application of the divergence theorem shows that the force and torque on
the surface of the swimmer are the same as the force and torque on the fluid at infinity. Using this
equivalence, the components of the force on the swimmer are
\beqa
	F_i &=& \lim_{R\to\infty}\int_0^{2\pi} (T_{ij}r_j) R d\vphi
\eeqa
and the torque on the swimmer is
\beqa
	N = \lim_{R\to\infty}\int_0^{2\pi} (\epsilon_{ij} r_i T_{jk} r_k) R^2 d\vphi \ .
\eeqa
In these formulas $r_i$ are the components of the radial unit vector 
$\hat{\mb{r}} = \cos\vphi\hat{\mb{x}} + \sin\vphi\hat{\mb{y}}$ and 
the integral is taken over the circle at infinity. 

Using these equations, and the components of the odd-viscosity part of the stress tensor, we can derive expressions
for the odd-viscosity contribution to the force and torque on the swimmer. In complex form they are
\beq
	F^o = \lim_{R\to\infty} -2\eta^o \oint_{\mathcal{C}} (\partial_{\bar{z}} v)\ d\bar{z} 
\eeq
and
\beq
	N^o = \lim_{R\to\infty} -2\eta^o\ \text{Re}\left\{ i\oint_{\mathcal{C}} z (\partial_{z}\bar{v})\ dz \right\}
\eeq
where $\mathcal{C}$ is a circular contour of radius $R$ (to be taken to infinity), and we have switched to a complex notation
for the force, $F = F_1 + iF_2$ (the torque, being a scalar in 2D, is real). 

Plugging in the velocity expansion \eqref{velocity} into these formulas gives
\begin{subequations}
\beqa
	F^o &=& 0 \\
	N^o &=& -4\pi\eta^o\text{Re}[b_{-2}] \label{odd_torque}\ .
\eeqa
\end{subequations}
In the next subsection we will show that the physical interpretation of this result is that the odd-viscosity contribution
to the torque is proportional to the flux of the fluid at infinity (see Section \ref{sec:PhysInt}). 
Previously it has been shown \cite{Wilc1989} that the even-viscosity contribution to the force and torque on the 
swimmer is given by
\begin{subequations}
\beqa
	F^e &=& 0 \\
	N^e &=& 4\pi\eta^e\text{Im}[b_{-2}] \label{even_torque}\ .
\eeqa
\end{subequations}
The swimmer feels no net force (a generic result for Stokes flows in two dimensions \cite{Avr2004}) and the total
torque is
\beq
	N = 4\pi( \eta^e \text{Im}[b_{-2}] - \eta^o\text{Re}[b_{-2}])\ . \label{total_torque}
\eeq

We can cancel the torque on the swimmer by having the swimmer rotate at a certain angular velocity $\omega$. This 
uniquely determines the rotational part of the gauge potential. In dimensions $D>2$ the translational part of the
gauge potential can be determined by the condition that the net force on the swimmer vanish. In two dimensions, however,
the net force vanishes identically \cite{Avr2004} and so one must instead determine 
the translational part of the gauge potential by requiring
that the fluid velocity vanish at infinity \cite{Acheson}. We discuss this condition in more detail in Section \ref{sec:Motion}.

\subsection{Physical Interpretation of the Torque Formula}
\label{sec:PhysInt}

The physical content of the formula \eqref{total_torque} for the net torque on the swimmer can be better understood by
looking at the relation of the coefficient $b_{-2}$ to the circulation and flux of the fluid at infinity, denoted by $\Gamma(\infty)$ 
and $\Phi(\infty)$, respectively. We can express the circulation and flux of the fluid at infinity in the form of line integrals
of the velocity around a large circle of radius $R$, to be taken to infinity. We have,
\beq
	\Gamma(\infty) = \lim_{R\to\infty}\int_0^{2\pi} \mb{v}\cdot\hat{\mbs{\vphi}} R d\vphi
\eeq
and
\beq
	\Phi(\infty) = \lim_{R\to\infty}\int_0^{2\pi} \mb{v}\cdot\hat{\mb{r}} R d\vphi \label{flux_infty}\ .
\eeq
Using the velocity expansion \eqref{velocity}, we find
\begin{subequations}
\beqa
	\Gamma(\infty) &=& -2\pi\text{Im}[b_{-2}] \\
	\Phi(\infty) &=& 2\pi\text{Re}[b_{-2}]\ .
\eeqa
\end{subequations}
Using these expressions, the net torque on the swimmer can be rewritten in the form
\beq
	N = -2\eta^e\Gamma(\infty) -2\eta^o\Phi(\infty)\ .
\eeq
The condition of vanishing torque in the different cases can then be interpreted in terms of zero circulation at infinity for even
viscosity only, zero flux at infinity for odd viscosity only, or a proportionality between the flux and circulation at infinity when
both types of viscosity are present.

\section{Model Swimming Strokes and Area Formula}

Following Ref. \onlinecite{Wilc1989}, we will begin by considering nearly circular swimmers with swimming strokes of the form
\beq
	S_0(\sg,t)= \al_0(t)\sg + \al_{-2}(t)\sg^{-1} + \al_{-3}(t)\sg^{-2} \label{stroke} \ ,
\eeq
where the $\al_{i}(t)$'s are coefficients which determine the time evolution of the 
swimming stroke. This kind of stroke is just a conformal map of degree ${\cal{D}}=2$ from the unit circle to the
complex $z$-plane. The absence of a term $\al_{-1}(t)$ ``fixes the gauge" with respect to translations \cite{Wilc1989}.
An important formula is the area of the swimmer at time $t$, which is given by
\beqa
	A(t) &=& \frac{1}{2}\text{Im}\left\{\oint \overline{S_0(\th,t)}\ dS_0(\th,t)\right\} \nnb \\
                  &=& \frac{1}{2}\text{Im}\left\{\int_0^{2\pi} \overline{S_0(\th,t)}\ \frac{dS_0(\th,t)}{d\th} d\th \right\} \label{area}\ ,
\eeqa
which gives
\beq
	A(t)= \pi(|\al_{0}|^2 - |\al_{-2}|^2 - 2|\al_{-3}|^2)
\eeq
for the simple stroke \eqref{stroke}. 
General swimmers represented by conformal maps of degree ${\cal{D}}$ have the form
\beq
	S_0(\sg,t)= \al_0(t)\sg + \sum_{n=1}^{{\cal{D}}} \al_{-n}(t) \sg^{-n} \label{gen-stroke}
\eeq
and in Appendix \ref{app:Degree3} we extend the swimming motion formulae to swimmers with ${\cal{D}}=3$.

\section{Solution for Translational and Rotational Motion of Swimmer}
\label{sec:Motion}

To determine the coefficients $a_k$ and $b_k$ in the velocity expansion \eqref{velocity} we need to conformally map the 
flow field back to the $\zeta = re^{i\theta}$ plane \cite{Wilc1989}. Recall that the shape of the swimmer $S_0(\sg,t)$
is a conformal map in the other direction, from the unit circle $\sg = e^{i\th}$ in the $\zeta$-plane to the $z$-plane. For general
swimmers of the form \eqref{gen-stroke} the conformal mappings between the $\zeta$ and $z$ planes take 
the form \cite{Wilc1989}, 
\begin{subequations}
\label{CM}
\beqa
	z &=& S_0(\zeta) = \al_0(t)\zeta + \sum_{n=1}^{\cal{D}} \al_{-n}(t) \zeta^{-n} \label{CT1}\\
	\zeta &=&  S_0^{-1}(z) = \frac{z}{\al_0} - \frac{\al_{-2}}{z} + \dots\ .
\eeqa
\end{subequations}
We now introduce a star $*$ symbol to denote the pull-back of a function in the $z$-plane to the $\zeta$-plane obtained by 
substituting \eqref{CT1} for $z$ in that function. The pull-backs of $\phi_1(z)$ and $\phi_2(z)$ are denoted by 
\begin{subequations}
\beqa
	\phi_1^*(\zeta) &=& \sum_{k<0}a_k^* \zeta^{k+1} \\
	\phi_2^*(\zeta) &=& \sum_{k<-1}b_k^* \zeta^{k+1} \ ,
\eeqa
\end{subequations} 
where the $a_k^*$ and $b_k^*$ are a new set of coefficients related to the original $a_k$ and $b_k$ through
the conformal mapping. 

Next we pull back the velocity field onto the unit circle $\sg$ in the $\zeta$-plane so that we can apply the no-slip
boundary conditions there and determine the pull-back coefficients $a_k^*$ and $b_k^*$ in terms of the $\al_{i}(t)$.
On the unit circle $\sg$ the velocity expansion takes the form (suppressing the $t$ dependence)
\beq
	v^*(\sg) = \phi_1^*(\sg) - \frac{S(\sg)}{\overline{\psg S(\sg)}}\overline{\psg\phi_1^{*} (\sg)} + \overline{\phi_2^*(\sg)}\ . \label{velocity_pb}
\eeq

The only coefficients we need to determine the translational and rotational motion of the swimmer are $a_{-1}$ and
$b_{-2}$. This is because $a_{-1}$ gives the fluid flow at infinity, so it determines the translational motion of the swimmer,
and $b_{-2}$ is related to the torque on the swimmer, so it determines the rotational motion of the swimmer. Using the 
conformal mapping \eqref{CM}, the coefficients $a_{-1}$ and $b_{-2}$ can be expressed in terms of the pulled-back
coefficients $a_k^*$ and $b_k^*$ as
\begin{subequations}
\beqa
	a_{-1} &=& a_{-1}^{*} \\
	b_{-2} &=& \al_{0}b_{-2}^{*}\ .
\eeqa
\end{subequations}
We can solve for the pulled-back coefficients $a_k^*$ and $b_k^*$ in terms of the parameters $\al_{i}$ using 
\eqref{velocity_pb}, and then use the pulled-back coefficients to solve for $a_{-1}$ and $b_{-2}$.
As in \cite{Wilc1989} we find
\begin{subequations}
\beqa
	a_{-1} &=& -\bar{\al}_0^{-1}\al_{-3}\dot{\bar{\al}}_{-2} \\
	b_{-2} &=& \bar{\al}_0\dot{\al}_0 - \al_{-2}\dot{\bar{\al}}_{-2} - 2\al_{-3}\dot{\bar{\al}}_{-3}\ . \label{b-2}
\eeqa
\end{subequations}

To determine the translational part of the gauge potential we note that
the coefficient $a_{-1}$ is a constant contribution to the velocity expansion, which means that the fluid velocity at 
infinity is uniform and non-zero. Following Section 7.5 of Ref.~\onlinecite{Acheson}, 
we argue that a finite-size swimmer located near the origin
should not be able to induce a non-zero fluid velocity at infinity, and so we make a Galilean transformation
to a frame in which the fluid is at rest at infinity and the swimmer moves with a velocity
\beq
	\A_{tr} \equiv V_x + iV_y = -a_{-1}\ ,
\eeq
where $\A_{tr}$ denotes the translational part of the gauge potential \eqref{gauge}.

To determine the rotational part of the gauge potential we attempt to cancel the torque \eqref{total_torque} on 
the swimmer by having the swimmer rotate at an appropriately chosen angular velocity $\omega$. In the 
parameterization \eqref{stroke} of the swimming stroke, having the swimmer rotate at an angular velocity $\omega$
amounts to the replacement 
\beq
	\al_{i} \to \al_{i,rot} = \al_i e^{i\omega t}\ .
\eeq
Under this replacement we find
\beq
	b_{-2} \to b_{-2,rot} = b_{-2} +  i\omega \left( |\al_0|^2 + |\al_{-2}|^2 + 2|\al_{-3}|^2 \right)\ ,
\eeq
so that the condition that the net torque on the swimmer vanish becomes
\beq
	\eta^e \text{Im}[b_{-2,rot}] - \eta^o\text{Re}[b_{-2,rot}] = 0\ .
\eeq
Solving this equation for $\omega$ yields the rotational part of the gauge potential
\beq
	\A_{rot} \equiv \omega =  \frac{-\text{Im}[b_{-2}] +  \tfrac{\eta^o}{\eta^e}\text{Re}[b_{-2}]}{|\al_0|^2 + |\al_{-2}|^2 + 2|\al_{-3}|^2 }\ . \label{A_rot}
\eeq

This expression shows that in the presence of odd viscosity the rotational part of the gauge potential picks up 
a term proportional to $\text{Re}[b_{-2}]$.
For the simple swimming stroke \eqref{stroke}, one can verify by explicit computation that 
\beq
	\text{Re}[b_{-2}] = \frac{1}{2\pi}\frac{dA(t)}{dt} \label{area_relation}\ ,
\eeq
which shows that the odd viscosity contribution to the angular velocity of the swimmer is proportional to the rate of change
of the area of the swimmer. This conclusion is not limited to swimming strokes which are conformal maps of degree ${\cal{D}}=2$, but
 holds for \emph{generic} swimmers bounded by a closed curve without any self-intersections, as we prove in Appendix
\ref{sec:AreaProof}.

We can use this area relation and the relation $\Gamma(\infty) = -2\pi\text{Im}[b_{-2}]$ for the circulation of the fluid 
at infinity to rewrite the angular velocity
formula in a way which clearly shows the physical meaning of each term. We find
\beq
	 \omega =  \frac{\Gamma(\infty)  +  \tfrac{\eta^o}{\eta^e}\frac{dA(t)}{dt}}{2\pi(|\al_0|^2 + |\al_{-2}|^2 + 2|\al_{-3}|^2) }\ .
\eeq

As the ratio of $\eta^o/\eta^e$ increases, the angular velocity \eqref{A_rot} grows without bound. Therefore we conclude
that in a fluid in which the odd viscosity terms completely dominate the stress tensor (i.e. $\eta^o/\eta^e \to \infty$), 
the condition that the swimmer experience zero net 
torque must be satisfied by taking $\frac{dA(t)}{dt} = 0$, otherwise the angular velocity of the swimmer would have to be
infinite. So a swimmer in a fluid where odd viscosity effects are dominant must have constant area. 

When discussing the limit
$\eta^o/\eta^e \to \infty$ in this context, we must always assume that $\eta^e$ is finite and large enough so that we can
still neglect any inertial forces in the problem (and so we can still take advantage of the geometric formulation of the problem
of swimming at low Reynolds number). This is why we have been careful to say ``when the odd viscosity is dominant" and not
``when $\eta^e = 0$." 

\section{Example Swimming Strokes}
\begin{figure}[t]
  \centering
    \includegraphics[width= .5\textwidth]{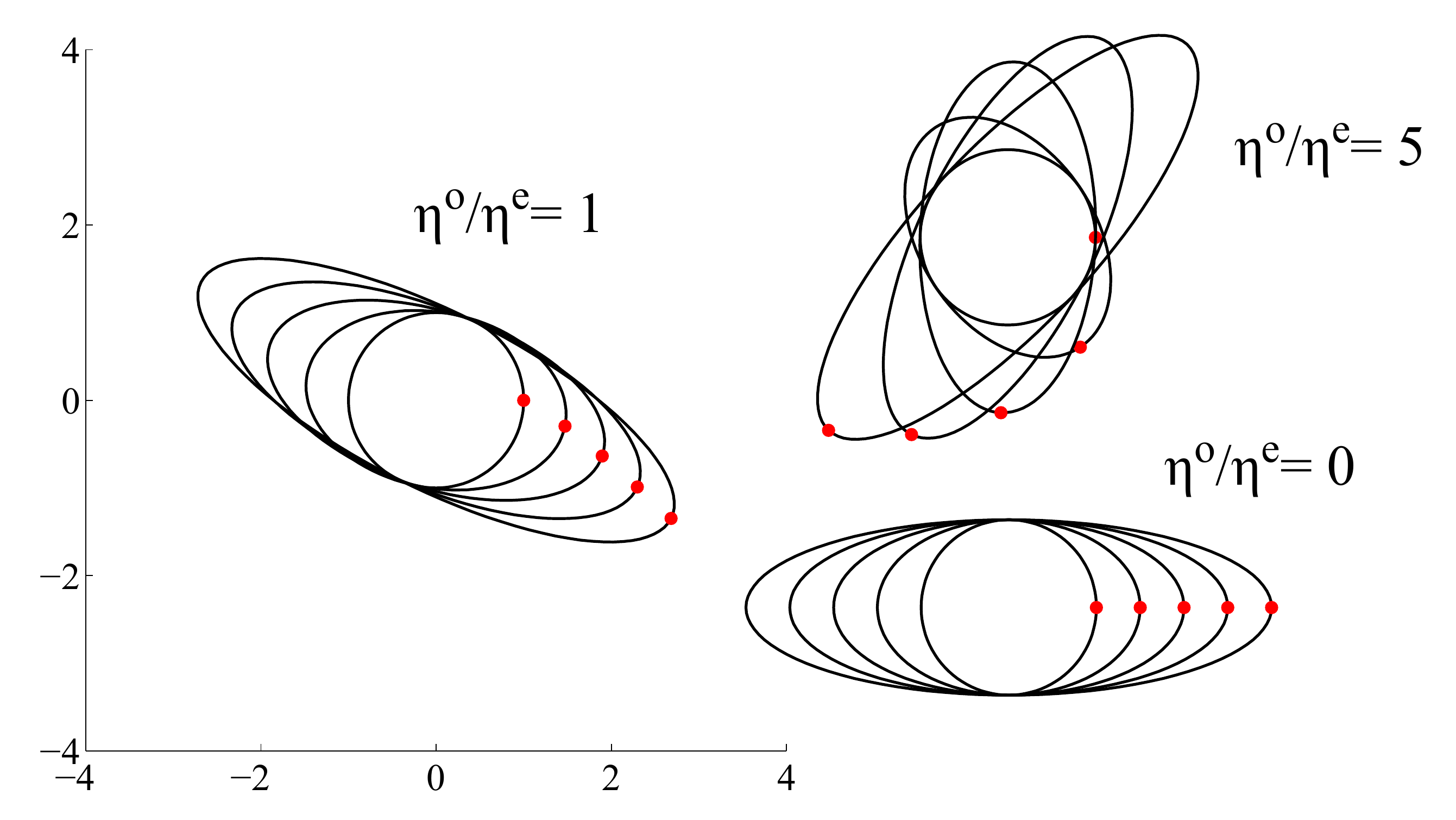} 
\vskip 10pt
 \caption{The elliptical distortion given by Eq. \eqref{ellipse}, shown in three different fluids with different ratios
of odd to even viscosity. The time between each consecutive shape is $0.5$ units of time. 
 The red dot is a guide for the eye that indicates the same point on the boundary of the shape, so one can clearly see
when the shape is rotating and when it is stationary.}
\label{fig:ellipse}
\end{figure}
Here we present some simple examples of swimming strokes that clearly demonstrate the difference between
swimming in a fluid with just even viscosity and swimming in a fluid with both even and odd viscosity.
\subsubsection{Dipolar Distortion}
The first example is a swimmer which starts out as a circle but grows into an ellipse by elongating one of its axes through 
a dipolar-like distortion. We use the parameterization
\begin{subequations}
\label{ellipse}
\beqa
	\al_0 &=& 1+\tfrac{t}{2} \\
	\al_{-2} &=& \tfrac{t}{2}  \\
	\al_{-3} &=& 0
\eeqa
\end{subequations}\noindent for this swimmer. 
The boundary of the swimmer is an ellipse with the lengths of the major and minor axes given by $a= 1+t$, $b=1$.
With only the conventional even viscosity this stroke will not cause any motion other than an increase in the area. We can also see this from the reflection symmetry about the x-axis,
which is equivalent to the fact that all the coefficients are real. However, when there is also odd viscosity this swimmer will start to rotate 
because its area is growing and the torque has a term proportional to the odd viscosity and the rate of area change. The motion for different values of the odd viscosity can be seen in Fig. \ref{fig:ellipse}.

\subsubsection{Quadrupolar Distortion}
To further test our results we chose a swimmer with a more complicated quadrupolar distortion which also has a uniform area growth. We used the parameterization
\begin{subequations}
\label{dwave}
\beqa
	\al_0 &=& 1+t \\
	\al_{-2} &=& 0 \\
	\al_{-3} &=& 0 \\
	\al_{-4} &=& \tfrac{1}{4}\ .
\eeqa
\end{subequations}
This swimming parameterization represents a conformal map of degree ${\cal{D}}=3.$ To see how to extend the analysis of the 
previous section to swimmers which are
conformal maps of the circle of degree ${\cal{D}}=3$ (i.e. how to include $\al_{-4}$
terms), see Appendix \ref{app:Degree3}. In Fig. \ref{fig:dwave} we see very similar results to the dipolar case, e.g., the motion 
of the swimmer is just a rotation proportional to the growth of the area. This indicates, as we expected from the general result
of Appendix \ref{sec:AreaProof}, 
that the odd viscosity does not distinguish between different types of shape distortions, and only couples to changes in the total 
area of the interior of the swimmer. 

\begin{figure}[t]
 \centering
    \includegraphics[width= .5\textwidth]{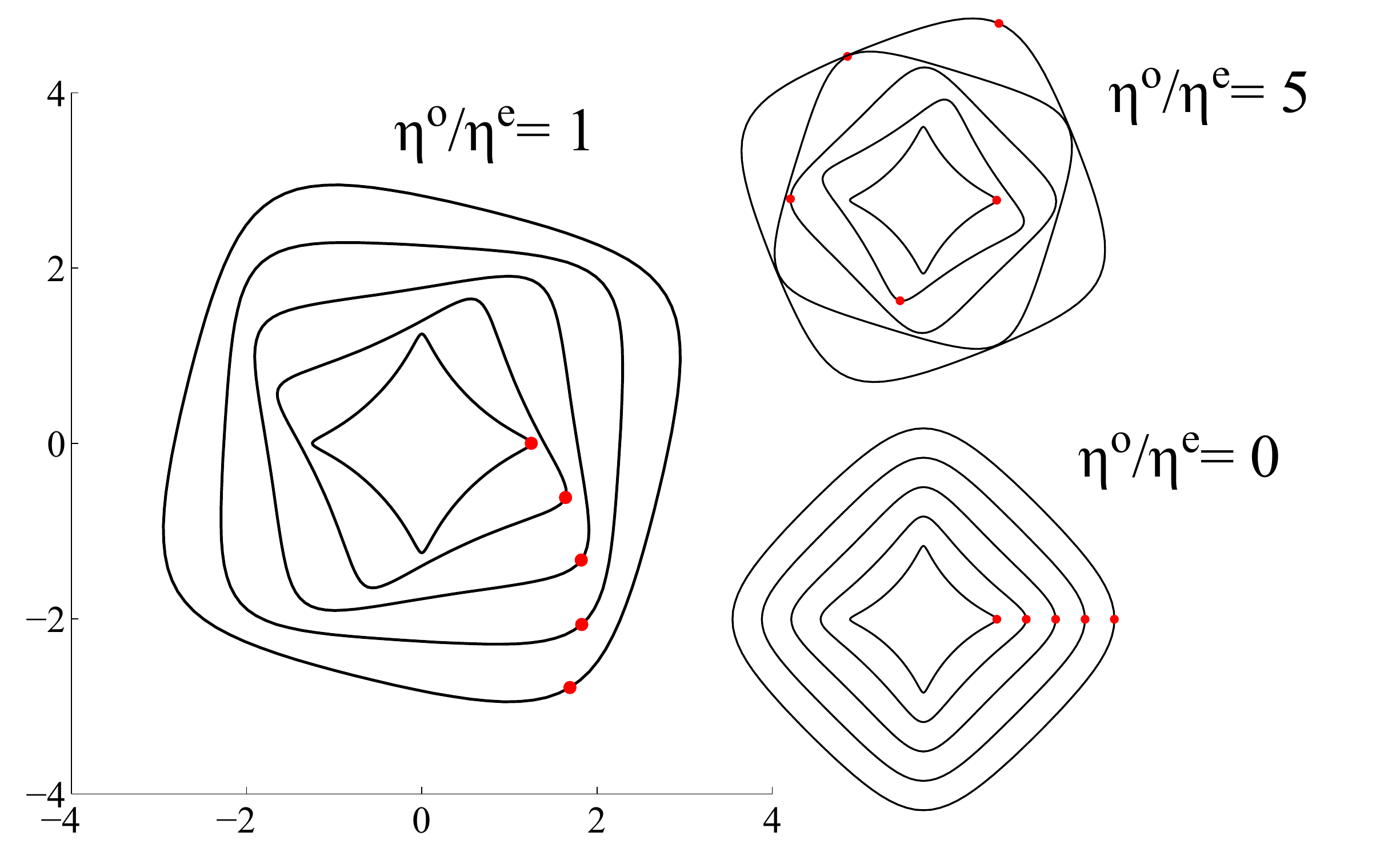}
\vskip 10pt
  \caption{The quadrupolar distortion given by Eq. \eqref{dwave}, shown in three different fluids with different ratios
of odd to even viscosity. The time between each consecutive shape is $0.5$ units of time. 
 The red dot is a guide for the eye that indicates the same point on the boundary of the shape, so one can clearly see
when the shape is rotating and when it is not.}
\label{fig:dwave}
\end{figure}
\begin{figure}[t]
 \centering
 \includegraphics[width= .5\textwidth]{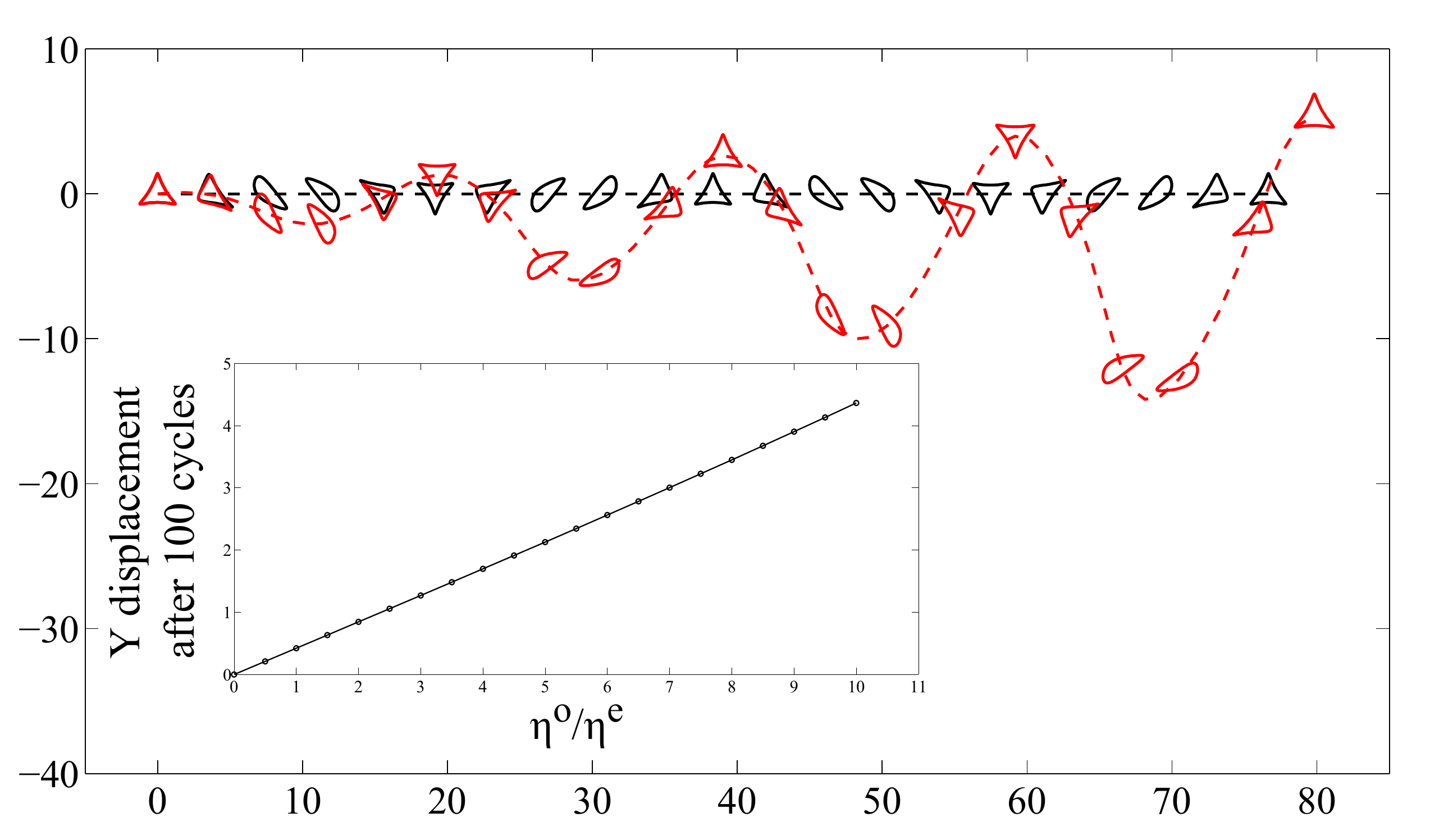}
\vskip 10pt
  \caption{The swimming stroke of Eq. \eqref{wandering} with the parameter values $r_0=1$, $\xi_1= 0.5$, and $\xi_2= 0.4$, shown
first with just even viscosity (in black) and then with both even and odd viscosity (in red) with $\eta^o/\eta^e = 10$. 
The time between each consecutive shape in the figure is $6.1$ cycles. 
 When odd viscosity is also present, the swimmer wanders off of its straight trajectory
because of rotations caused by changes in the area of the swimmer. The inset shows the y-displacement of the
swimmer after 100 cycles of this swimming stroke vs. the ratio of the odd and even viscosity coefficients.}
\label{fig:wandering_trajectory}
\end{figure}
\subsubsection{Wandering Stroke}
The third example is a swimmer parameterized with the cyclic stroke
\begin{subequations}
\label{wandering}
\beqa
	\al_0 &=& r_0 \\
	\al_{-2} &=& -i\xi_1 \sin(2\pi t) \\
	\al_{-3} &=& -i\xi_2 \cos(2\pi t)
\eeqa
\end{subequations}
where $r_0$, $\xi_1$ and $\xi_2$ are all real parameters. We chose this particular stroke because in the case when only the even viscosity is present, the swimmer's centroid moves in a straight line through the fluid. Additionally, this stroke has a periodic time-dependent area
\beq
	A(t) = \pi[ r_0^2 - \xi_1^2 + (\xi_1^2 - 2\xi_2^2)\cos^2(2\pi t) ]\ ,
\eeq\noindent which implies that it will feel a cyclic stress from the odd viscosity term when present. In Fig. \ref{fig:wandering_trajectory} one can clearly see the outcome as we show two trajectories, one with $\eta^o/\eta^e=0$ and one with $\eta^o/\eta^e=10.$ In the case when $\eta^o$ vanishes, the swimmer travels in a straight line, however in the second case the swimmer oscillates transverse to the straight-line path. In the inset we show that the amplitude of the transverse oscillation at a fixed time increases linearly with the slope $\eta^o/\eta^e.$ As the swimmer continues it will wander further and further off of the straight-line course although on average it seems like it will still progress linearly at a similar rate to that of the swimmer in the fluid with vanishing odd viscosity.

\subsubsection{Null-Rotation Stroke}
The fourth example is a stroke which will nominally rotate when just even viscosity is present, but for which the variation of the area of the shape
has been chosen carefully so that when odd viscosity is also present the shape will not rotate at all. 
In other words, the odd viscosity contribution to the angular velocity exactly cancels the even viscosity contribution for
a given particular ratio $\eta^o/\eta^e$ which, for the sake of this example, we pick to be unity.

A glance at Eq. \eqref{A_rot} shows that in order to produce this cancellation, we need the stroke to satisfy
\beq
	\text{Im}[b_{-2}] = \text{Re}[b_{-2}]\ .
\eeq
For swimmers which are conformal maps of the circle of degree ${\cal{D}}=2$, the coefficient $b_{-2}$ is given by Eq. 
\eqref{b-2}. We see from that equation that we can design such a stroke by taking 
$\al_0 = r_0 = \text{constant}$ and 
\beq
	\al_j(t) = r_j(t) e^{i\theta_j(t)} \label{cancel-rot}
\eeq
for $j= -2,-3$, where the functions $r_j(t)$ and $\theta_j(t)$ are functions which are determined in the following way.
We would like to have 
\beq
	\al_j(t)\dot{\bar{\al}}_j(t) = (1+i)\dot{f}_j(t) \label{cancel-2}
\eeq
where the $f_j(t)$ are some real periodic functions of time (to give a periodic swimming stroke), which we are essentially
free to choose. 
This choice will guarantee the cancellation of the even and odd viscosity contributions to the torque on the swimmer, since
the real and imaginary parts of Eq. \eqref{cancel-2} are equal at all times.
The reason for using the
derivative of the functions $f_j(t)$ in the above formula is purely for convenience in the formulas that follow. 
Plugging the form \eqref{cancel-rot} for the $\al_j(t)$ into this last equation and solving the two coupled ordinary differential
equations for $r_j(t)$ and $\theta_j(t)$ gives the form of the stroke in terms of the functions $f_j(t)$,
\begin{subequations}
\beqa
	r_j(t) &=& \sqrt{2\big(f_j(t) + C_{j,1}\big)} \\
	\theta_j(t) &=& -\frac{1}{2}\ln\big(f_j(t) + C_{j,1}\big) + C_{j,2}\ ,
\eeqa 
\end{subequations}
where $C_{j,1}$ and $C_{j,2}$ are arbitrary constants (although they must be chosen carefully along with the functions 
$f_j$ to keep the argument of the logarithm from ever equaling zero). Now any choice of the periodic functions $f_j(t)$ will 
give a cyclic swimming stroke that will not rotate in a fluid with our chosen ratio $\eta^o/\eta^e=1$.

This shows in principle that it is possible to construct a stroke for which the even and odd viscosity 
contributions to the angular velocity
exactly cancel each other. Swimmers using this type of stroke might be able to more efficiently navigate odd-viscosity fluids since 
the particular choice of stroke cancels the rotation effects due to the odd-viscosity. 

\section{Conclusion}

We have applied the geometric theory of swimming at low Reynolds number
developed by Wilczek and Shapere~\cite{Wilc1989} to the case where the fluid
has a non-vanishing Hall, or odd, viscosity. The main effect of the Hall viscosity is to introduce an additional torque on the
swimmer, proportional to the rate of change of the area of the swimmer, independent of the other shape changes occurring in the stroke pattern. This torque is the companion effect to the fact that
a swimmer rotating in a fluid with odd viscosity feels an inwards or outwards pressure proportional to its 
angular velocity~\cite{Avr1998}.
As we show in Appendix \ref{sec:AreaProof} this
conclusion applies to generic swimming shapes and is not limited to swimmers whose boundaries are simple
conformal maps of the unit circle. 

As a consequence of this extra torque, a swimming stroke which would not cause the swimmer to rotate in a fluid with 
conventional viscosity can cause the swimmer to rotate in a fluid with Hall viscosity if the area of the swimmer is changing. 
It is even possible to design a stroke which will rotate the swimmer in an even viscosity fluid but not in a fluid
with both even and odd viscosity, for a certain value of the ratio $\eta^o/\eta^e$. It is possible that swimmers placed in fluids with an odd viscosity would have to adapt their strokes to efficiently move in a straight line. Additionally, it would be interesting to see if swimmers could use the interplay between the even and odd viscosity to perform more interesting or efficient motion patterns.

\begin{acknowledgements}
We acknowledge useful discussions with J. E. Avron and P. Zhang. TLH acknowledges support from DOE QMN under grant 
number DEFG02-07ER46453. We are thankful for the support of the Institute for Condensed Matter Theory at UIUC.

\end{acknowledgements}

\appendix
\section{Computation of the Path-Ordered Integral}
\label{app:computation}

To calculate the matrix $\R(t)$, which gives the rigid motion of the swimmer after a finite time $t$, we need to
evaluate the reverse path-ordered integral \eqref{POI}. In practice we do this by slicing time into many 
small steps (say N steps) of size $\Delta t$. We can write 
\beq
	\bar{P}e^{\int_0^{t} \A(t') dt'} = \bar{P}e^{\sum_{i=1}^N \int_{(i-1)\Delta t}^{i\Delta t} \A(t') dt'}\ .
\eeq
If the time steps are small enough then we can approximate this as
\beq
\bar{P}e^{\sum_{i=1}^N \int_{(i-1)\Delta t}^{i\Delta t} \A(t') dt'} \approx \prod_{i=1}^N \bar{P}e^{\int_{(i-1)\Delta t}^{i\Delta t} \A(t') dt'}\ ,
\eeq
where on the right side we now have a product of reverse path-ordered integrals over many small time intervals 
of size $\Delta t$ and we should put the earliest times on the right so that we are applying the rigid motions in these small
intervals in chronological order. Since these time intervals are very small we can make a further approximation by expanding the
path-ordered integral over the time interval $\Delta t$ to first order and neglecting the higher order terms to find:
\beq
\bar{P}e^{\int_{(i-1)\Delta t}^{i\Delta t} \A(t') dt'} \approx I + \int_{(i-1)\Delta t}^{i\Delta t} \A(t') dt' \ . 
\eeq
Finally we can make one further approximation for the integral of the matrix $\A(t)$ over the small time interval $\Delta t$,
\beq
	\int_{t_{i-1}}^{t_i} \A(t') dt' \approx \A(t_{i-1})\Delta t
\eeq
where $t_i = i\Delta t$ (and $t_0 = 0$). Our final expression for the approximation of the full path-ordered integral 
is then
\beq
	\bar{P}e^{\int_0^{t} \A(t') dt'} \approx \prod_{i=1}^N \left( I +  \A(t_{i-1})\Delta t \right)\ ,
\eeq
where again the matrices for the earliest times must to be to the right so that the rigid motions are applied in the proper order.

We have also tried expanding the reverse path-ordered integrals over the time interval $\Delta t$ to second order, but
it seems that this makes almost no visible correction to the swimming trajectory when the swimming deformations are not
too large and the step size $\Delta t$ is small.

\section{Proof that $\text{Re}[b_{-2}] = \frac{1}{2\pi}\frac{dA(t)}{dt}$ for general swimming strokes}
\label{sec:AreaProof}

Our analysis of the simple swimmer \eqref{stroke} suggests a deeper connection between the area of the swimmer
and the odd viscosity contribution to the torque on the swimmer. To explore this connection further we now show that
Eq. \eqref{area_relation} holds for any swimmer whose boundary is a smooth curve without self-intersections.

The boundary of the swimmer is just a smooth curve parameterized by $\theta$ which also depends on the time $t$. If
we write the shape in terms of real components
\beq
	S_0(\theta,t) = x(\theta,t) + iy(\theta,t)
\eeq
and plug into the area formula \eqref{area} we find
\beq
	A(t) = \frac{1}{2}\int_{0}^{2\pi} \left[x(\theta,t)y'(\theta,t) - y(\theta,t)x'(\theta,t)  \right]d\theta\ ,
\eeq
where the prime denotes a derivative with respect to $\theta$. Next take a time derivative to get
\beqa
	\frac{dA(t)}{dt} &=& \frac{1}{2}\int_{0}^{2\pi} \left[\dot{x}(\theta,t)y'(\theta,t) + x(\theta,t)\dot{y}'(\theta,t) \right. \nnb \\ & &  - \left. \dot{y}(\theta,t)x'(\theta,t) -  y(\theta,t)\dot{x}'(\theta,t) \right]d\theta\ .
\eeqa
We can integrate by parts on the terms with mixed partial derivatives and use the fact that the boundary terms
vanish since $x(\theta,t)$, $y(\theta,t)$, $\dot{x}(\theta,t)$ and $\dot{y}(\theta,t)$ are $2\pi$-periodic in $\th$ to 
get
\beq
	\frac{dA(t)}{dt} =  \int_{0}^{2\pi} \left[\dot{x}(\theta,t)y'(\theta,t) - \dot{y}(\theta,t)x'(\theta,t)\right]d\theta\ .
\eeq 
Because of the no-slip boundary conditions the vector $(\dot{x}(\theta,t),\dot{y}(\theta,t))$ is just the fluid velocity 
$\mb{v}(\mb{r})$ evaluated on the surface of the swimmer,
\beq
	\mb{v}(\mb{r})|_{\text{swimmer}} = \dot{x}(\theta,t)\hat{\mb{x}} + \dot{y}(\theta,t)\hat{\mb{y}}\ .
\eeq
Then we can write 
\beq
	\frac{dA(t)}{dt} = \oint_{\text{swimmer}} \mb{v}\cdot\hat{\mb{n}}\ ds = \Phi(\text{swimmer})\ ,
\eeq
where $\hat{\mb{n}}\ ds = \hat{\mb{z}}\times d\mb{r}$ is a vector normal to the surface of the swimmer
 with magnitude $ds = |d\mb{r}|$. This integral is just the flux of the fluid at the surface of the swimmer. 
By the divergence theorem we have
\beq
	\Phi(\infty) - \Phi(\text{swimmer}) = \int_{\text{fluid}} \nabla\cdot\mb{v}\ dxdy
\eeq
and since the fluid is incompressible, $\nabla\cdot\mb{v} = 0$, we get
\beq
	\frac{dA(t)}{dt} = \Phi(\infty)\ .
\eeq
A comparison with Eq. \eqref{flux_infty} for the flux of the fluid at infinity yields the final result
\beq
	\text{Re}[b_{-2}] = \frac{1}{2\pi}\frac{dA(t)}{dt}\ ,
\eeq
proving that this relation is valid for general swimming shapes in incompressible fluids.

It is known that an object which rotates in a fluid with odd viscosity will feel a pressure directed radially inwards or
outwards depending on the direction of the rotation \cite{Avr1998}. The relation \eqref{area_relation} is the companion
to this statement. It says that an object which tries to expand or contract in a fluid with odd viscosity will feel a torque
whose direction ($\pm \hat{\mb{z}}$) depends on whether the area of the object is growing or shrinking.

\section{Extension and solution of conformal maps of degree ${\cal{D}}=3$}
\label{app:Degree3}

A swimmer whose boundary is a degree 3 (${\cal{D}}=3$) conformal map of the circle has the form
\beq
	S_0(\sg,t)= \al_0(t)\sg + \al_{-2}(t)\sg^{-1} + \al_{-3}(t)\sg^{-2} + \al_{-4}\sg^{-3} \ ,
\eeq
with area
\beq
	A(t)= \pi(|\al_{0}|^2 - |\al_{-2}|^2 - 2|\al_{-3}|^2 - 3|\al_{-4}|^2)\ .
\eeq

To solve for $a_{-1}^*$ and $b_{-2}^*$ we need
the coefficients $a_{-2}^{*}, a_{-3}^{*}$ and $a_{-4}^{*}$. Equations for these coefficients can be obtained
by plugging into the pulled-back velocity expansion \eqref{velocity_pb}. We find that
\begin{subequations}
\beqa
	\al_{-4}\bar{a}_{-2}^{*} + \bar{\al_0}a_{-2}^{*} &=& \bar{\al_0}\dot{\al}_{-2} \\
	a_{-3}^{*} &=& \dot{\al}_{-3} \\
	a_{-4}^{*} &=& \dot{\al}_{-4}\ .
\eeqa 
\end{subequations}
The equation for $a_{-2}^{*}$ is really just a matrix equation for a two-component vector consisting of the real and
imaginary parts of $a_{-2}^{*}$. The solution is
\beq
	a_{-2}^{*} = \frac{|\al_0|^2\dot{\al}_{-2} - \al_0\al_{-4}\dot{\bar{\al}}_{-2}}{|\al_0|^2-|\al_{-4}|^2}\ .
\eeq
In terms of this coefficient we find that 
\beq
	a_{-1} = -(\bar{\al}_0)^{-1}(\bar{a}_{-2}^{*}\al_{-3} + 2\dot{\al}_{-3}\al_{-4})
\eeq
and
\beqa
	b_{-2} &=& \bar{\al}_0\dot{\al}_0 - \bar{\al}_{-2}\dot{\al}_{-2} - 2\al_{-3}\dot{\bar{\al}}_{-3} \nnb \\ &-& 3\al_{-4}\dot{\bar{\al}}_{-4} + \bar{\al}_{-2}a_{-2}^{*} - \al_{-2}\bar{a}_{-2}^{*}\ . 
\eeqa
Note that the last two terms in $b_{-2}$ are complex conjugates of each other and appear with the opposite
sign so that they will cancel when we take the real part of $b_{-2}$. This means that the relation 
 $\text{Re}[b_{-2}] = \frac{1}{2\pi}\frac{dA(t)}{dt}$ still holds in this case, 
as we expect based on the general arguments presented in Appendix \ref{sec:AreaProof}.

To solve for the new form of the angular velocity necessary to cancel the torque on the swimmer, we again send
$\al_{i} \to \al_{i,rot} = \al_i e^{i\omega t}$ and solve the equation
\beq
	\eta^e \text{Im}[b_{-2,rot}] - \eta^o\text{Re}[b_{-2,rot}] = 0\ ,
\eeq
where now
\beq
	b_{-2,rot} = b_{-2} + i\omega J
\eeq
with 
\beqa
	J &=& \frac{2}{|\al_0|^2-|\al_{-4}|^2}\left(|\al_0|^2|\al_{-2}|^2 + \text{Re}[\al_0\al_{-4}(\bar{\al}_{-2})^2]  \right) \nnb \\ 
&+& |\al_0|^2 - |\al_{-2}|^2 + 2|\al_{-3}|^2 + 3|\al_{-4}|^2 \ .
\eeqa
The new angular velocity needed to cancel the torque on the swimmer is then
\beq
	\omega = \frac{1}{J}(-\text{Im}[b_{-2}] +  \tfrac{\eta^o}{\eta^e}\text{Re}[b_{-2}])\ ,
\eeq
so that the translational and rotational parts of the gauge potential are now given by
\beqa
	\A_{tr} &=& (\bar{\al}_0)^{-1}(\bar{a}_{-2}^{*}\al_{-3} + 2\dot{\al}_{-3}\al_{-4}) \\
	\A_{rot} &=& \frac{1}{J}(-\text{Im}[b_{-2}] +  \tfrac{\eta^o}{\eta^e}\text{Re}[b_{-2}])\ .
\eeqa

\end{document}